# Introduction d'un modèle d'engrènement original et effet sur le comportement modal d'une transmission

# (Effect of an original gear mesh modelling on the modal behaviour of gear transmission)

Emmanuel RIGAUD[a], Joël PERRET-LIAUDET[a], Mohamed-Salah MECIBAH[b]

(a) Laboratoire de Tribologie et Dynamique des Systèmes,

UMR 5513 CNRS/ECL/ENISE

Ecole Centrale de Lyon. 36, avenue Guy de Collongue

69134 ECULLY cedex, France

(b) Département de Génie-Mécanique,

Faculté des Sciences de l'Ingénieur

Université de Constantine

Constantine, Algérie

Correspondance :

e-mail : emmanuel.rigaud@ec-lyon.fr

# Introduction d'un modèle d'engrènement original et effet sur le comportement dynamique d'une transmission (Effect of an original gear mesh modelling on the gearbox dynamic behaviour)

**Résumé**

La prédiction du comportement vibroacoustique des transmissions s'appuient sur la détermination des sources excitatrices, la génération de modèles globaux à grands nombres de degrés de liberté, l'analyse modale et la résolution des équations paramétriques du mouvement générées à partir de ces modèles. Dans la mise en œuvre de tels modèles le couplage élastique entre les roues dentées induit par l'engrènement est le plus souvent décrit par la raideur d'engrènement. Néanmoins, lorsque le contact s'établit sur toute la largeur de la denture, le moment résultant des efforts transmis peut contraindre les angles de rotation de chaque roue dentée associés aux basculements des roues et aux mouvements de flexion des arbres. L'objectif de cette étude est d'introduire les termes de couplage entre roues prenant en compte ce phénomène. On montre également sur la base de quelques exemples dans quelles mesures ils peuvent influencer le comportement modal de la transmission et sa réponse vibroacoustique.

**Summary**

Prediction of the vibroacoustic behaviour of geared transmissions is generally based on determining the excitation sources, the overall modelling of the transmission, the modal analysis and the solving of the parametric equations of the motion generated by these models. On the building process of such large degrees of freedom models, the elastic coupling induced by the gear mesh is generally described by the meshing stiffness. Nevertheless, when the contact is established along the tooth width, the resulting moment can constrain the rotational angles associated to wheel tilting and flexural deformation of shafts. The scope of this study is to introduce the coupling terms between wheels associated to this phenomenum. One shows also on the basis of some examples in what manner they can influence the modal behaviour and the vibroacoustic response of the transmission.

# 1. Introduction

Le comportement vibroacoustique des boîtes de vitesses et, plus généralement, des transmissions par engrenages a pour origine principale l'erreur statistique de transmission [1-3]. Celle-ci constitue une source d'excitation interne, générée au niveau de l'engrènement, et dont les origines physiques sont les déformations élastiques des dentures dues à l'application du couple moteur et les écarts de géométrie de l'engrenage (défauts de fabrication, corrections de denture, erreurs de parallélisme induites par les défauts de positionnement des lignes d'arbres ou les déformations élastiques des roulements et du carter).

En régime de fonctionnement stationnaire, ces sources d'excitation sont à l'origine d'une fluctuation périodique de la raideur d'engrènement, raideur qui résulte du contact entre les différents couples de dents en prise. Elles génèrent des surcharges dynamiques sur les dentures qui sont transmises aux lignes d'arbres, aux roulements et au carter de la transmission. Les vibrations qui résultent de l'excitation de ces différentes composantes sont à l'origine de nuisances acoustiques. C'est l'état vibratoire du carter de la transmission qui constitue la principale source du bruit rayonné [3]. Dans ce cadre, nous nous intéressons aux modèles dynamiques globaux permettant de prédire ce comportement vibroacoustique.

# 2. Démarche de modélisation globale des transmissions par engrenages

La prédiction du comportement vibroacoustique des transmissions s'appuie en premier lieu sur la mise en œuvre d'outils permettant de caractériser la source excitatrice, à partir d'une modélisation des déformations élastiques des dentures et des écarts de géométrie des engrenages, et de la résolution des équations de contact entre dents en prise [4-6].

Elle nécessite en second lieu la mise en œuvre de modélisation globale des transmissions par la génération de modèles éléments finis à grand nombre de degrés de liberté, et la résolution d'équations de mouvement paramétriques générées à partir de ces modèles.

Les méthodes de discrétisation des différents éléments qui composent une transmissions par engrenages (roues dentées, lignes d'arbres, accouplements, moteur, charge, carter) s'appuient sur la méthode des éléments finis et sont décrites par exemple dans [7-9]. Pour la prise en compte des roulements, des matrices de raideur généralisées sont généralement introduites. Elles sont déduites d'un calcul d'équilibre statique préalable [10,11].

A titre d'illustration, la figure 1 présente un maillage de lignes d'arbres et des roulements d'une transmission simple étage à axes parallèles.

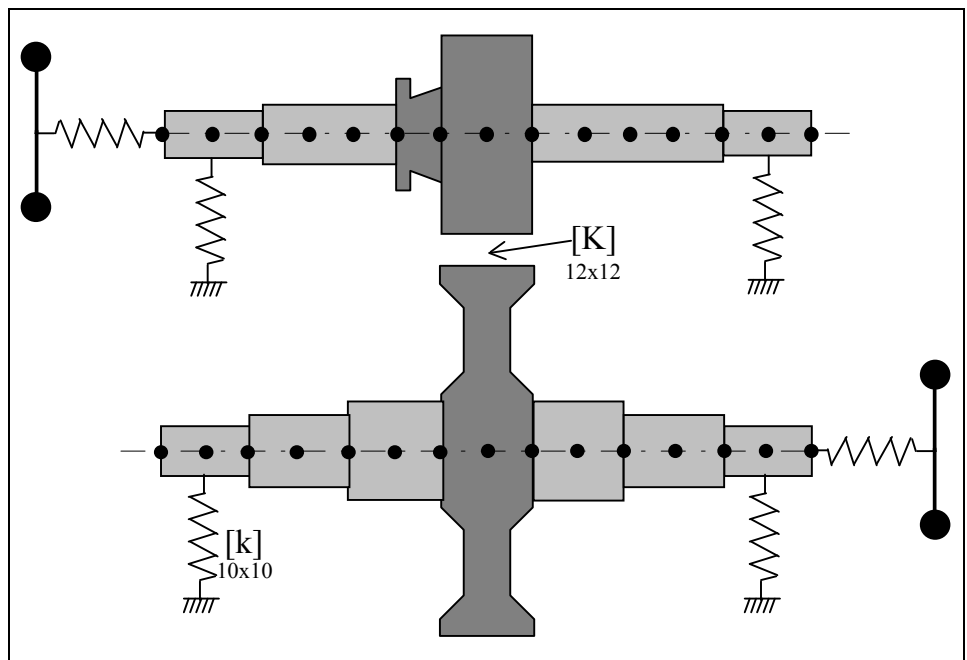

*Figure 1. Maillage des lignes d'arbres et des roulements.*

Typiquement, il n'est pas rare de traiter des modèles de transmissions réelles comportant de l'ordre de 100 000 degrés de liberté, permettant d'intégrer les propriétés élastiques du carter. Ces modèles globaux sont nécessaires car, d'une part, plusieurs travaux montrent clairement que les conditions instantanées de l'engrènement sont fortement influencées par le comportement dynamique des autres composants de la transmission (tels que les arbres, les roulements et le carter) et, d'autre part, ces modèles globaux sont les seuls qui permettent d'accéder à la réponse vibratoire normale de l'enveloppe qui constitue la source de rayonnement acoustique de la transmission [7,12]. Par contre, de tels modèles ne permettent pas de décrire finement les actions de contact au niveau des engrènements qui sont le plus souvent décrits par une raideur dite raideur d'engrènement, linéarisée autour de l'équilibre statique et projetée sur le plan d'action théorique selon la normale aux dentures. La prise en compte de la moyenne temporelle de la raideur d'engrènement dans les modèles globaux permet alors de déterminer les caractéristiques modales de la transmission.

**3. Modélisation des caractéristiques élastiques de l'engrènement**

Les caractéristiques élastiques du couplage induit par l'engrènement sont déduites par linéarisation des efforts de contact autour de l'équilibre statique. Afin d'introduire ces caractéristiques dans les modèles dynamiques, il convient donc de résoudre les équations de l'équilibre statique du contact entre dents.

**3.1 Equations du contact et répartition de charge**

Les équations d'équilibre reposent sur l'hypothèse généralement admise que les efforts entre dents restent normaux aux profils théoriques et sont donc contenus dans le plan d'action. Cette hypothèse suppose d'une part que l'on néglige les composantes dues au frottement, et d'autre part que les défauts de géométrie, comme les déformations élastiques, ont une influence négligeable sur la normale aux profils. Dans ce cadre, et pour une position angulaire donnée de l'arbre d'entrée $\Theta_1$, les équations du contact s'écrivent sous la forme suivante [5] :

$$\int_\Gamma G_f(x,s)p(s)ds + G_h(x)p(x) = [D(x)-e(x)].H_C(x) \qquad (1)$$

$$\int_\Gamma p(s)ds.\cos\beta = \frac{C_1}{Rb_1} = \frac{C_2}{Rb_2} = F \qquad (2)$$

Dans cette équation et pour la position angulaire choisie $\Theta_1$, $\Gamma$ constitue l'ensemble des lignes théoriques de contact, $p(s)$ constitue la distribution linéique de la charge transmise, $G_f(x,s)$ constitue un terme de compliance couplant le déplacement normal en $x$ induit par une charge unitaire en $s$, $G_h(x)$ constitue un terme de compliance associé à l'écrasement local de matière (contact étroit), $D(x)$ représente le rapprochement normal des flancs théoriques de denture induit par les mouvements de corps rigide des deux roues, $e(x)$ permet de modéliser les défauts de géométrie (écart initial selon la normale aux profils entre les flancs de denture), $H_C(x)$ est la fonction échelon prenant la valeur 1 si la quantité ($D(x)-e(x)$) est positive, 0 si elle est négative, $\beta$ constitue l'angle d'hélice de base, $C_1$ et $C_2$ sont les couples extérieurs appliqués au pignon et à la roue, $Rb_1$ et $Rb_2$ les rayons de base et $F$ la force normale à la denture transmise par l'engrenage. La résolution de cette équation permet d'évaluer les deux grandeurs inconnues que sont la répartition linéique de l'effort $p(s)$ et le rapprochement local entre dents induit par les mouvements de corps rigide des roues en prise $D(x)$. En principe, la méthode numérique mise en œuvre est basée sur la discrétisation spatiale des équations (1) et (2) le long des lignes de contact théorique, puis sur l'emploi d'une technique de résolution telle que par exemple la méthode du simplexe [7,13]. Les caractéristiques de compliance sont en général déduites de modélisation préliminaire des dentures, par exemple par la méthode des éléments finis. A partir de ce calcul statique, il est possible de déduire les caractéristiques élastiques du couplage induit par l'engrènement. On se propose de revenir en premier lieu sur la description de la démarche classique, puis de l'étendre à celle proposée dans cet article.

**3.2 Modèle classique du couplage entre les roues dentées en prise**

Les éléments de réduction de la distribution des efforts linéiques $p(x)$ se réduisent au point primitif $I$ défini dans le plan médian des roues à la force normale à la denture $F$, et à un moment $M$ porté par l'axe perpendiculaire au plan d'action (voir *figure 2*). La force $F$ est décrite par l'équation (2) et le moment $M$ s'exprime aisément sous la forme suivante :

$$M = \int_\Gamma L(s) p(s) ds \qquad (3)$$

Dans cette équation, la fonction $L(s)$ constitue la distance entre le pied de la perpendiculaire à la ligne de contact passant par le point primitif $I$, et le point de contact $Q(s)$ considéré sur cette ligne (voir *figure 2*). Par ailleurs, et compte tenu de l'hypothèse des petits déplacements, le rapprochement local $D(s)$ en chaque point $Q(s)$ du contact théorique peut être modélisé par un champ de déplacement dont les éléments de réduction sont le rapprochement normal à la denture $d_n$ et une rotation $\lambda$ autour de l'axe perpendiculaire au plan d'action passant par le point primitif $I$ de telle sorte que :

$$D(s) = d_n + L(s).\lambda \qquad (4)$$

Les modèles usuels du couplage élastique sont basés sur l'hypothèse selon laquelle les effets du moment $M$ et de l'angle $\lambda$ sont négligeables. Dans ce cas, on définit le couplage élastique en introduisant un simple ressort de raideur $k$ (raideur d'engrènement) agissant selon la normale au profil et passant par le point primitif $I$. Pratiquement, on accède à cette grandeur en estimant la dérivée de la force $F$ par rapport au rapprochement $d_n$ :

$$k = \frac{\partial F}{\partial d_n} \approx \frac{F(d_n + \delta d_n) - F(d_n)}{\delta d_n} \approx \frac{\delta F}{d_n(F + \delta F) - d_n(F)} \quad (5)$$

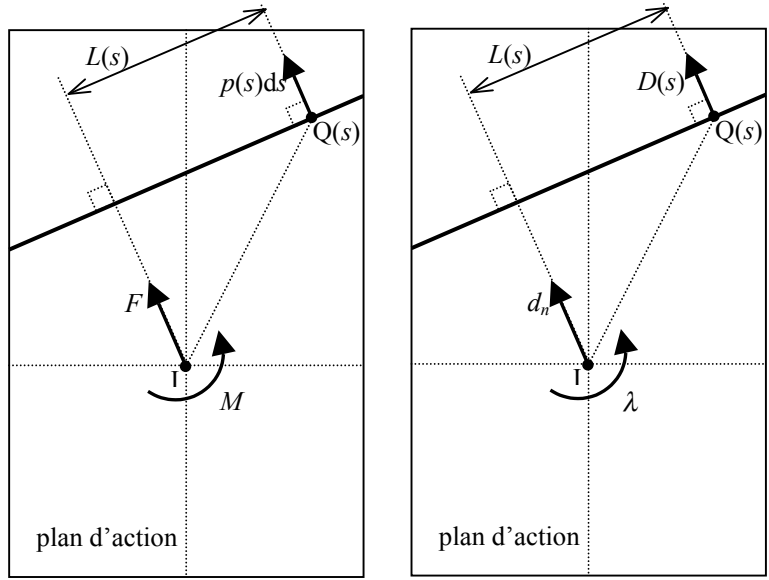

*Figure 2 : Description des efforts et des déplacements sur le plan d'action*

La raideur d'engrènement dépend de la position angulaire de la roue d'entrée $\Theta_1$ et en régime stationnaire, elle peut être assimilée à une fonction périodique du temps $k(t)$ de fréquence égale à la fréquence d'engrènement. La matrice de raideur généralisée de rang 12 couplant les 6 degrés de liberté du nœud central de la roue menante aux 6 degrés de liberté du nœud central de la roue menée peut être déduite du calcul de l'énergie potentielle élastique :

$$U_P = \frac{1}{2} d_n k d_n \quad (6)$$

En introduisant la relation qui relie les 12 degrés de liberté précédents (vecteur des coordonnées généralisées **q**) au rapprochement $d_n$ :

$$d_n = {}^t\mathbf{T}.\mathbf{q} \quad (7)$$

On en déduit la matrice de raideur généralisée :

$$\mathbf{K} = k(t)\mathbf{T}.^t\mathbf{T} \qquad (8)$$

A titre d'exemple, dans le repère local orthonormé direct (**X**, **Y**, **Z**) défini *figure 3*

$$^t\mathbf{T} = < 0, \cos\beta, \sin\beta, Rb_1 \sin\beta \tan\alpha, -Rb_1 \sin\beta, Rb_1,$$
$$0, -\cos\beta, -\sin\beta, Rb_2 \sin\beta \tan\alpha, -Rb_2 \sin\beta, Rb_2 > \qquad (9)$$

où $\alpha$ est l'angle de pression apparent de fonctionnement.

Enfin, pour évaluer les caractéristiques modales de la transmission, on introduit la moyenne temporelle de la raideur d'engrènement $k_m$ en lieu et place de $k(t)$ dans l'équation (8).

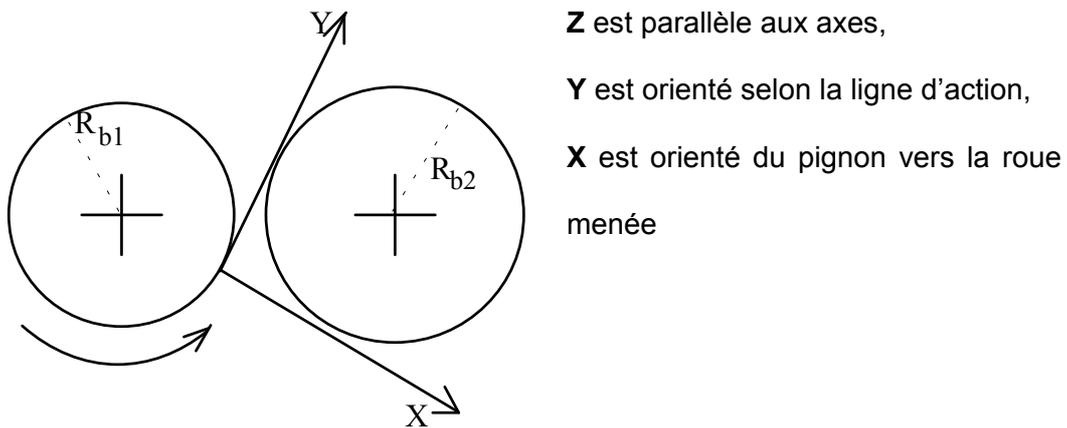

**Z** est parallèle aux axes,

**Y** est orienté selon la ligne d'action,

**X** est orienté du pignon vers la roue menée

*Figure 3. Plan apparent de l'engrenage. Définition du repère local.*

### 3.3 Introduction d'un modèle original du couplage élastique

Le principal défaut associé au modèle précédent est qu'il ne permet pas de décrire les effets longitudinaux de la répartition de charge sur les dents et donc les caractéristiques de raideur associées au moment $M$ et à l'angle de rotation $\lambda$. Pourtant ces effets se traduisent a priori par des contraintes sur les rotations des sections libres des arbres $\theta_{x1}$ et $\theta_{x2}$ dans le cas d'engrenages droits mais aussi $\theta_{y1}$ et $\theta_{y2}$ si la denture est hélicoïdale. L'objectif de cette étude est d'introduire ces effets et de montrer dans quelles mesures ils peuvent influencer le comportement modal d'une transmission et donc sa réponse vibroacoustique. Pour ce faire, on linéarise les expressions de la force normale $F$ et du moment $M$ vis-à-vis des déplacements généralisés $d_n$ et $\lambda$. Ainsi, on obtient formellement :

$$\begin{Bmatrix} F \\ M \end{Bmatrix} = \begin{bmatrix} k_{11} & k_{12} \\ k_{21} & k_{22} \end{bmatrix} \begin{Bmatrix} d_n \\ \lambda \end{Bmatrix} \qquad (10)$$

à partir des évaluations numériques de $k_{11}$, $k_{12}$, $k_{21}$ et $k_{22}$ définis par :

$$k_{11} = \frac{\partial F}{\partial d_n}, \quad k_{12} = \frac{\partial F}{\partial \lambda}, \quad k_{21} = \frac{\partial M}{\partial d_n}, \text{ et } \quad k_{22} = \frac{\partial M}{\partial \lambda} \qquad (11)$$

Dans une démarche similaire au cas précédent, il est possible d'introduire la matrice de raideur généralisée symétrique (de rang 12) couplant les 6 degrés de liberté du nœud central de la roue menante aux 6 degrés de liberté du nœud central de la roue menée. Il vient :

$$\mathbf{K} = \mathbf{R}\,\mathbf{k}(t)\,^{\mathrm{t}}\mathbf{R} \qquad (12)$$

où $\mathbf{k}(t)$ est la matrice de raideur de rang 2 de l'équation (10) et $\mathbf{R}$ est cette fois ci une matrice 12 x 2 dépendant des caractéristiques géométriques de l'engrenage et couplant les 12 degrés de liberté (vecteur $\mathbf{q}$) au vecteur constitué du rapprochement $d_n$ et de l'angle $\lambda$.

Afin d'illustrer simplement la démarche, nous nous proposons ci après de considérer le cas d'un engrenage droit pour lequel un modèle simplifié de type Winkler est introduit pour décrire l'élasticité de la denture.

### 3.4 Exemple

Dans le repère local décrit *figure 3*, il est aisé de montrer que la matrice $\mathbf{R}$ s'écrit :

$$^t\mathbf{R} = \begin{bmatrix} 0 & 1 & 0 & 0 & 0 & Rb_1 & 0 & -1 & 0 & 0 & 0 & Rb_2 \\ 0 & 0 & 0 & 1 & 0 & 0 & 0 & 0 & 0 & -1 & 0 & 0 \end{bmatrix} \quad (13)$$

Par ailleurs, dans le cas d'un modèle d'élasticité des dentures de type Winckler (raideur locale constante, équirépartie et indépendante en chaque point discret des lignes de contact), les termes $k_{11}$, $k_{12}$, $k_{21}$ et $k_{22}$ peuvent être calculées aisément connaissant la répartition de charge instantanée sur la denture. Ainsi, si la répartition est établie sur toute la largeur de denture la matrice de raideur $\mathbf{k}(t)$ s'écrit :

$$\mathbf{k}(t) = k_{11}(t) \begin{bmatrix} 1 & 0 \\ 0 & \dfrac{b^2}{12} \end{bmatrix} \quad (14)$$

où $b$ constitue la largeur de denture. Par contre, si la répartition est partielle, elle est distincte du cas précédent. Pour l'exemple d'une répartition établie sur la moitié de la largeur de denture (cas extrême qui résulterait d'un mésalignement important des arbres), $\mathbf{k}(t)$ s'écrit :

$$\mathbf{k}(t) = k_{11}(t) \begin{bmatrix} 1 & -\dfrac{b}{4} \\ -\dfrac{b}{4} & +\dfrac{b^2}{12} \end{bmatrix} \quad (15)$$

Remarquons que dans le premier cas (équation (14)), il n'existe pas de couplage entre $d_n$ et $\lambda$ au contraire du deuxième cas (équation (15)).

**4. Simulations numériques**

**4.1 Transmission étudiée**

Les résultats qui suivent correspondent à un engrenage parallèle 28/58 dents à denture droite. Ses caractéristiques géométriques principales sont un module de 2,5 mm et un angle de pression de 20°. La largeur de denture correspond à un paramètre variable compris entre 20 et 50 mm.

Cet engrenage est monté sur deux jeux de lignes d'arbres tels que les roues dentées sont soit centrées sur des arbres courts de longueur entre paliers égale à 100 mm, soit décentrées sur des arbres longs de longueur entre paliers égale à 180 mm.

Chaque ligne d'arbres est supportée par deux roulements à billes et reliée au moteur et à la charge par l'intermédiaire d'accouplements flexibles.

**4.2 Résultats**

Parmi l'ensemble des modes propres d'une transmission par engrenages, certains présentent une énergie de déformation importante localisée au niveau de la raideur d'engrènement $k$ (modes de denture). Les "vitesses critiques" de rotation, au sens ou l'on observe une amplification importante de la surcharge dynamique de denture et, plus généralement, de la réponse vibroacoustique de la transmission correspondent à l'excitation en résonance de ces modes de denture.

**4.2.1 Denture totalement chargée**

Lorsque l'engrenage est fortement chargé, le calcul de l'erreur statique de transmission montre que la charge se répartit sur toute la largeur de denture. Dans ce cas de figure, la raideur moyenne d'engrènement $k_{11}$ est égale à 300 N.µm$^{-1}$ et les valeurs moyennes des termes $k_{22}$ et $k_{12}$ sont fournies via l'équation (14).

Le calcul de la base modale de la transmission montre que les fréquences propres et les déformées, et notamment celles des modes de denture, sont très peu affectées par l'introduction du terme $k_{22}$ dans le modèle d'engrènement original, car l'amplitude de ce terme reste négligeable devant celle des autres termes de la matrice de raideur, même pour des engrenages très larges.

Par conséquent, la *figure 4* montre que l'évolution de l'effort dynamique de denture est très peu affectée, ce qui justifie a posteriori l'utilisation du modèle d'engrènement classique (paragraphe 3.2). Cette conclusion est valable aussi bien pour des roues dentées centrées sur des arbres courts (*figure 4a*) ou décentrées sur des arbres plus longs (*figure 4b*).

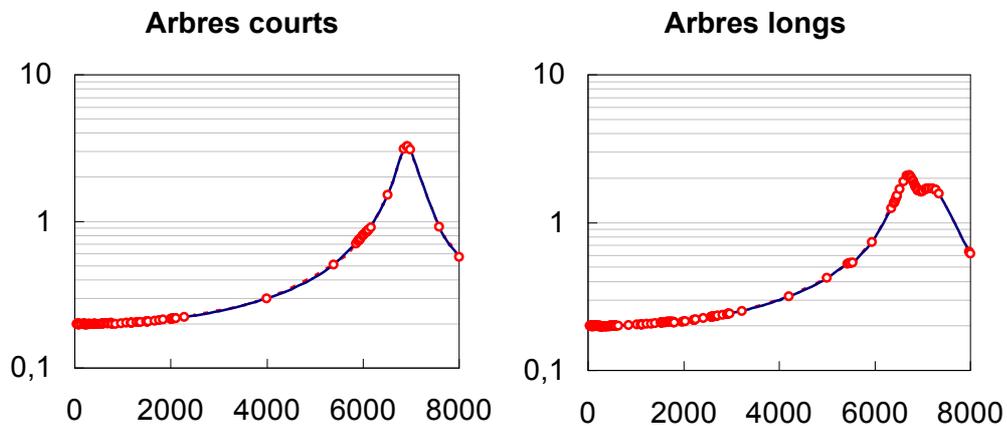

*Figure 4. Evolution de l'effort dynamique de denture (N) en fonction de la fréquence d'engrènement (Hz)*

*(modèle d'engrènement initial ∘∘∘∘∘∘, modèle d'engrènement original ——— ).*

*b=50 mm.*

Seul le calcul de la réponse dynamique de la transmission à une excitation paramétrique $k_{22}(t)$ permet de mettre en évidence une modification du comportement dynamique de la transmission. La *figure 5* montre que celle-ci présente de nouveaux pics d'amplification qui traduisent l'excitation de modes (autour de 500 et 4000 Hz) pour lesquels une part significative de l'énergie de déformation est stockée dans la raideur $k_{22}$.

Néanmoins, l'amplitude de la réponse de la transmission induite par cette fluctuation paramétrique $k_{22}(t)$ reste négligeable devant la réponse induite par la fluctuation paramétrique $k_{11}(t)$, que l'on considère la réponse de la denture, les efforts aux paliers ou la réponse du carter.

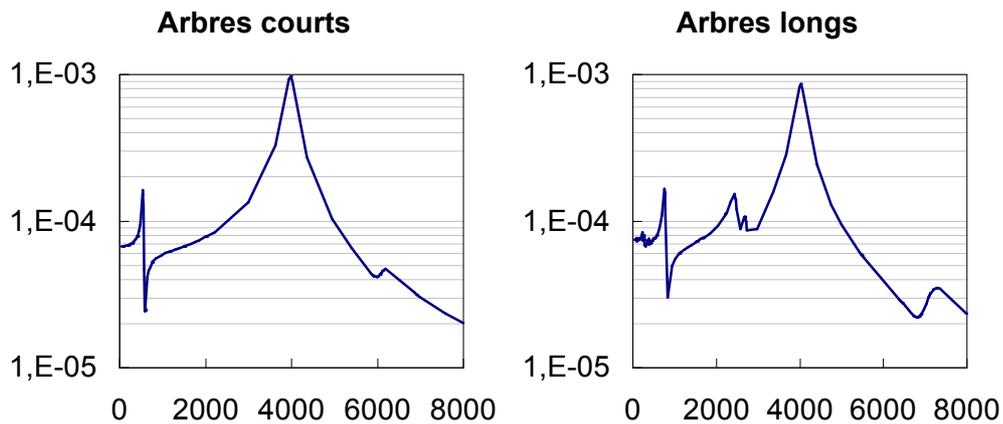

*Figure 5. Réponse dynamique de la denture à une excitation paramétrique $k_{22}(t)$. b=50 mm.*

**4.2.2 Denture partiellement chargée**

Lorsque l'effort statique appliqué sur l'engrenage est faible, le calcul de l'erreur statique de transmission montre que la charge se répartit sur une partie seulement de la denture. C'est le cas notamment si l'engrenage présente des défauts de parallélisme ou des défauts de denture longitudinaux significatifs. On observe alors une diminution significative de la raideur moyenne d'engrènement $k_{11}$.

En l'absence de défauts de parallélisme ou de défauts longitudinaux, la charge reste centrée par rapport au plan médian de l'engrenage et les valeurs moyennes des termes $k_{22}$ et $k_{12}$ sont déduites de l'équation (14) (notamment $k_{12}$=0). Les conclusions précédentes sont encore valables, ce qui signifie notamment que, dans ce cas de figure, le calcul de la réponse dynamique de la transmission à l'aide du modèle d'engrènement classique reste ici justifié.

Par contre, lorsque les défauts de parallélisme et les défauts longitudinaux sont importants, le calcul de l'erreur statique de transmission montre que la répartition de la charge sur la largeur de denture présente une forte dissymétrie.

On observe non seulement une diminution significative de la raideur moyenne d'engrènement $k_{11}$ (150 N.µm$^{-1}$), mais également une matrice symétrique de raideur pleine (équation (15)).

Les fréquences propres et les déformées des modes de denture initiaux sont légèrement affectées par l'introduction des termes $k_{12}$ et $k_{21}$ dans le modèle d'engrènement, si bien que les régimes de fonctionnement conduisant à une amplification maximale de la réponse dynamique ont augmenté, et ce d'autant plus que la denture est large.

D'autre part, la présence des termes croisés $k_{12}$ et $k_{21}$ induit un couplage entre les mouvements qui sollicitaient la raideur d'engrènement linéaire selon la normale au contact $k_{11}$ et les termes qui sollicitaient la raideur en rotation $k_{22}$. Certaines déformées modales sont telles qu'une partie de l'énergie de déformation initialement stockée dans la raideur $k_{22}$ est transférée dans $k_{11}$ si bien que ces modes peuvent être potentiellement excités par une fluctuation paramétrique $k_{22}(t)$ dont les effets restent négligeables, mais également par la fluctuation paramétrique dominante $k_{11}(t)$. La réponse dynamique de la transmission présente donc de nouveaux pics d'amplification significatifs non prédits avec le modèle d'engrènement initial. Les différentes simulations que nous avons effectuées montrent que la différence entre les réponses obtenues avec chacun des modèles est d'autant plus significative que la largeur de denture n'est plus négligeable devant la longueur des arbres ou que la denture est excentrée. Les *figures 6 et 7* illustrent ces résultats.

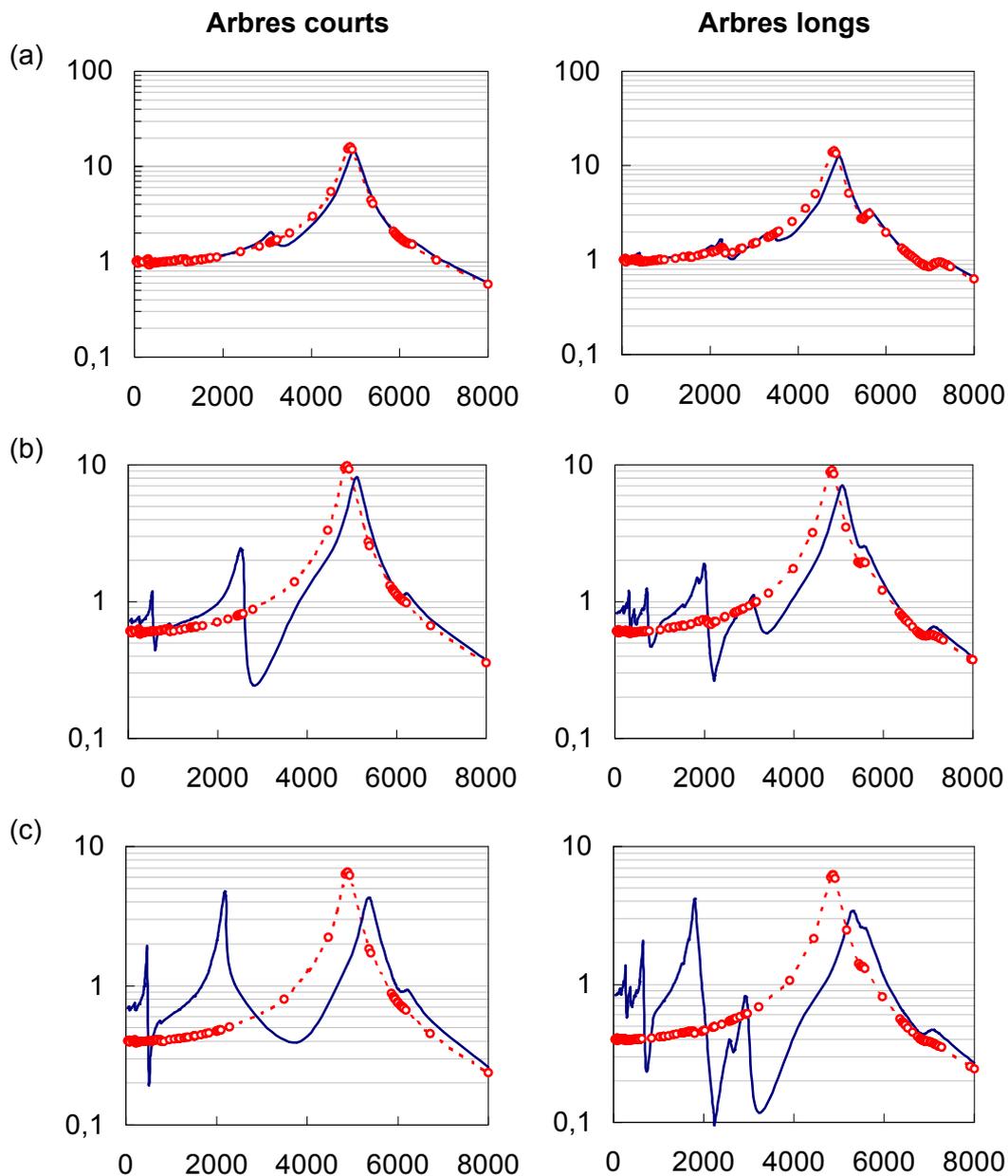

*Figure 6. Evolution de l'effort dynamique de denture (N) en fonction de la fréquence d'engrènement (Hz)*
*(modèle d'engrènement initial ∘∘∘∘∘∘∘, modèle d'engrènement original ——— ).*
*(a) b =20 mm (b) b =33 mm (c) b =50 mm*

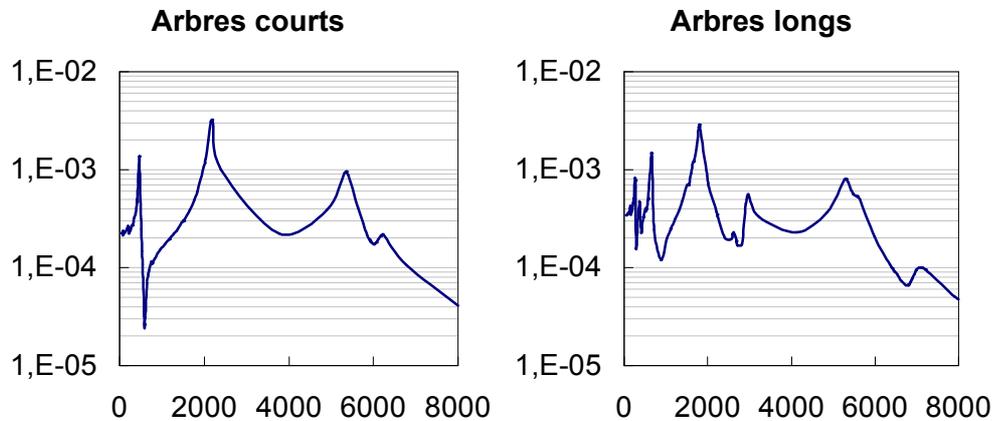

*Figure 7. Réponse dynamique de la denture à une excitation paramétrique $k_{22}(t)$. b=50 mm.*

## 5. Conclusion

Dans cette étude, le couplage élastique entre les roues dentées d'une transmission par engrenages est modélisé par une matrice de raideur qui prend en compte les contraintes sur certains angles de rotation de chaque roue dentée induites par le fait que le contact s'établit sur toute la largeur de la denture (basculement des roues). Ce modèle est substitué à la simple raideur d'engrènement projetée selon la normale au contact entre dentures. Bien que la modélisation classique s'avère justifiée dans la plupart des cas, les simulations effectuées ont montré que cette nouvelle modélisation était nécessaire pour décrire correctement le comportement dynamique d'une transmission dans certaines configurations particulières. Sous très faible charge et en présence de défauts de parallélisme et de défauts longitudinaux importants, la répartition des efforts sur la largeur de denture peut présenter une forte dissymétrie. Dans ce cas, dès lors que le ratio entre la largeur de denture et la longueur des arbres devient important, le couplage entre les mouvements des roues qui sollicitent l'engrènement selon la normale au contact et les mouvements de rotation des roues autour des axes autres que leur axe principal est tel que la réponse dynamique de la transmission présente de nouveaux pics d'amplification significatifs non prédits avec le modèle d'engrènement initial. Des travaux au sein de l'équipe sont en cours, pour préciser ces comportements d'une part dans le cadre de modèles de compliance plus réalistes et d'autre part dans le cadre des engrenages hélicoïdaux.

**6. Références**